\documentstyle[twocolumn,aps]{revtex}

\def\ket#1{|{#1}\rangle}

\def\swap{\begin{picture}(2,17)(2,0)
\put(-1,14){\framebox(2,2)}
\put(0,15){\line(0,-1){15}}
\put(-1,-1){\framebox(2,2)}
\end{picture}}

\def\h#1#2{\put(#1,#2){\framebox(10,10){$H$}}}
\def\c#1#2{\put(#1,#2){\framebox(10,10){$C$}}}
\def\bl#1{\line(1,0){#1}}
\def\bla#1#2#3{\put(#2,#3){\bl{#1}}}
\def\ml#1#2#3#4{\multiput(#1,#2)(0,15){#4}{\bl{#3}}}
\def\numa{\put(-7,73){4}
\put(-7,58){3}
\put(-7,43){2}
\put(-7,28){1}
\put(-7,13){0}}
\def\numb{\put(-7,73){4'}
\put(-7,58){3'}
\put(-7,43){2'}
\put(-7,28){1'}
\put(-7,13){0'}}
\def\numc{\put(-7,73){4''}
\put(-7,58){3''}
\put(-7,43){2''}
\put(-7,28){1''}
\put(-7,13){0''}}

\author{A. Saito\cite{A.S} \and
K. Kioi \and Y. Akagi \and N. Hashizume \and K. Ohta}
\title{Actual computational time-cost of the Quantum Fourier Transform 
in a quantum computer using nuclear spins}
\address{Advanced Technology Research Laboratories, Sharp Corporation,
Tenri-shi, 632 Japan}
\begin{document}
\maketitle

\begin{abstract}
There have been many proposed methods for the practical 
implementation of quantum computing. 
Now quantum computation has reached the turning point 
from being a conceptual system to becoming a physical one. 
In this paper, we discuss a practical elementary gate 
and the actual computational time-cost of the QFT 
in two physical implementations, 
namely the bulk spin resonance computer and
the Spin Resonance Transistor. 
We show that almost all universal gates require different times 
for operation.
The actual time-cost of the QFT is  $O(n2^n)$ for large $n$.
This differs drastically from the reported cost $O(n^2)$
based on ideal quantum computation.
\end{abstract}

Recent technological development has stimulated
proposals for many quantum computers:
Bulk Spin Resonance(BSR)\cite{gershenfeld,cory97,jones},
trapped ions\cite{cirac}, cavityQED\cite{turchette},
Josephson junctions\cite{averin},
coupled quantum dots\cite{Loss98} and
the Spin Resonance Transistor(SRT)\cite{Kane98,Vrijen99}.
BSR has been implemented experimentally
in some organic molecules by the use of conventional
nuclear magnetic resonance(NMR) equipment\cite{gershenfeld,cory97,jones}.
the SRT is attractive from the viewpoint of it's integration ability
and it's compatibility with silicon technology.

Complexity analysis classifies quantum algorithms according to
a function that describes how a computational cost
incurred in solving a problem scales up
as larger problems are considered\cite{Williams97}.
The computational cost of a quantum algorithm has usually been estimated
as the sum of the universal gates required in such ideal mathematical models as
the  Quantum Turing Machine(QTM) and the quantum circuit.
 The computational complexity 
is effective in estimating the essential performance of an algorithm
to factor out the variations in performance
experienced by different makes of computers 
with different amounts of Random Access Memory(RAM),
 swap space, and processor speeds.
The above cost is proportional to an actual time-cost 
in the physical implementation where all quantum operations 
can be achieved in the same time.
However, if the implementation being considered 
takes a different time for each quantum gate, 
there is a possibility that the actual time-cost 
will have a different behavior from the ideal cost.
A hardware dependent time-cost is important 
to experimentalists who research into practical implementations.
In this paper, we focus on the actual computational time-cost 
of the Quantum Fourier Transform (QFT) 
in two physical implementations that utilise nuclear spins:
BSR and the SRT.

For the following discussion,
we assumed that the quantum computers being considered 
are constructed from an array of n quantum registers 
labeled by $j(1 \leq j \leq n)$ in one dimension.
They calculate $n$ qubit data which corresponds to
$N=2^n$ states.
We defined the following matrices.
The suffixes on each matrix represent
the quantum registers on which it operates.
In the definitions of $C_{j,k}(\theta)$ and $D_{j,k}(\theta)$,
the first and the second suffixes represent the target bit  
and the controlled bit respectively.
\begin{displaymath}
H_{j} = \frac{1}{\sqrt{2}}
\left( \begin{array}{cc}
1 & 1 \\
1 & -1
\end{array}
\right)
, \ \
R_{yj}(\theta)=
\left( \begin{array}{cc}
\cos \theta/2  & \sin \theta/2  \\
- \sin \theta/2 & \cos \theta/2
\end{array}
\right)
\end{displaymath}

\begin{displaymath}
R_{zj}(\alpha) =
\left( \begin{array}{cc}
e^{i\alpha/2} & 0  \\
0 & e^{-i\alpha/2}
\end{array}
\right)
,\ \
C_{j,k}(\theta) =
\left( \begin{array}{cccc}
1 & 0 & 0 & 0 \\
0 & 1 & 0 & 0 \\
0 & 0 & 1 & 0 \\
0 & 0 & 0 & e^{i\theta}
\end{array}
\right)
\end{displaymath}

\begin{displaymath}
\Phi_{j}(\delta) =
\left( \begin{array}{cc}
e^{i\delta} & 0  \\
0 & e^{i\delta}
\end{array}
\right)
,
D_{j,k}(\theta) =
\left( \begin{array}{cccc}
e^{i\theta} & 0 & 0 & 0 \\
0 & e^{-i\theta} & 0 & 0 \\
0 & 0 & e^{-i\theta} & 0 \\
0 & 0 & 0 & e^{i\theta}
\end{array}
\right)
\end{displaymath}

Firstly, we discuss the practical elementary gate in these implementations.
The practical elementary gate is defined as 
a quantum gate that can be achieved directly 
by physical phenomena in the implementation being considered. 

BSR consists of quantum registers that are the nuclear spins of each atom
in an organic molecule\cite{gershenfeld,cory97,jones}.
External RF-pulses and magnetic fields control 
the quantum states and the quantum correlations in the following way.

The nuclear spin under a strong magnetic field ${\bf B} =(0,0,B_0)$
is described by the Hamiltonian
$H = - \gamma \hbar B_0 I_{jz}$,
where $\gamma$ is the gyro-magnetic ratio for the spin,
and $I_{jz}$ is  the z component of the $j$th nuclear spin.
The time evolution of the system
is $\exp( iHt/\hbar) = \exp(-i\gamma \hbar B_0 t S_{jz})$
and this then corresponds to the rotation on the z-axis
$R_{zj}(\theta=\gamma \hbar B_0 t)$.
The RF-pulse enables the rotation on the other axis\cite{Ernst94}.
The time evolution shows that the rotation angle $\theta$
can be controlled by two factors:
the intensity $B_0$, and the duration $t$,
of the external magnetic field.
Consequently, there are two control modes.
In the intensity control mode,
each single qubit rotation takes the same time.
In the duration control mode however, 
each single qubit phase rotation takes a different time,
which is proportional to the rotation angle.

The exchange interaction between the $j$-th and the $k$-th registers 
is described by
the  Hamiltonian $H_{exch.} = J_{jk} I_{jz} I_{kz}$,
where $J_{jk}$ is the exchange coupling constant.
The time evolution of the system is described by
\begin{equation}
\label{evolution}
\exp( i J_{jk} t I_{jz} I_{kz} / \hbar) = D_{j,k}(\theta = J_{jk} t / 2\hbar).
\end{equation}
Sequence(\ref{evolution}) shows that
the exchange interaction can control the phase rotation angle.
Any external fields, however, cannot control the interaction directly,
because atoms in a molecule  always interact with each other.
The refocusing technique can control
the angle $\theta$ effectively\cite{Ernst94}.
Only the time duration between the refocusing pulses 
determines the angle $\theta$.
The operation $D_{j,k}(\theta)$, therefore, takes a time 
that is proportional to the rotation angle  $\theta$.
An actual operation such as $D_{j,k}$ is effective 
only for adjacent registers
because the exchange interaction $J_{j,k}$ 
between non-adjacent registers is very small.

The SRT is composed of quantum registers which
are the nuclear spins of arrayed phosphorus ions in silicon, 
with a globally static magnetic field $B$ and an AC
magnetic field $B_{AC}$\cite{Kane98}.
The implementation consists of two gates on the surface:
the A-gate above each ion and the J-gate between adjacent ions.

The A-gate controls the strength of hyperfine interactions
and the resonance frequency of the nuclear spin beneath it.
A globally applied magnetic field $B_{AC}$
flips nuclear spins resonant with the field
by the same process that occurs in BSR.
In this case, only the duration of the resonance determines
the rotation angle.
It implies that each single qubit phase rotation always 
takes a different time.

The electron wave function extends over a large distance
and makes an effective electron-mediated coupling
for two nuclear spins sharing it in semiconductors.
The J-gate controls the overlap of the electron wave functions
bounded to two adjacent phosphorus atoms,
and hence  the electron-mediated exchange coupling $J_{j,k} = J(t)$ 
in the time evolution(\ref{evolution}) directly.
It therefore controls the phase rotation angle 
in the operation $D_{j,k}$.
The operation $D_{j,k}$ in this implementation also 
operates only for adjacent registers.

In both implementations,
all single qubit phase rotations and 
controlled phase rotations are practical elementary gates.
They can make the quantum XOR in the sequence below,
ordered from right to left\cite{gershenfeld}  
\begin{eqnarray} 
 & & \sqrt{-i} XOR(j,k) \nonumber \\
& = & R_{yj}(-\frac{\pi}{2}) R_{zk}(-\frac{\pi}{2}) 
R_{zj}(-\frac{\pi}{2}) D_{j,k}(\frac{\pi}{4})R_{yj}
(\frac{\pi}{2}).\label{xor_exchange}
\end{eqnarray}
The suffixes $j$ and $k$ represent the target bit and
the controlled bit respectively.
The sequence(\ref{xor_exchange}) shows
that the quantum XOR  depends on single qubit gates
in this technique, 
and is then concerned with 
the time required for phase rotation.

Barenco and co-workers showed that
other universal gates can be constructed by 
all single qubit gates and the quantum XOR\cite{Barenco95}.
Each $n(\geq 1)$ qubit gate required 
a different number of them\cite{Barenco95}, 
and then a different time for execution. 
As discussed above,
almost all universal gates take a different time 
in these implementations.
It suggests the  possibility that
the actual time-cost of the quantum algorithm is
different from the ideal  cost,
though the complexity is not affected in the QFT case.

Next we estimate the actual time-cost of the QFT
that could be achieved by the above practical elementary gates,
by considering the time resolutions of 
the controlling external fields 
in these implementations.

The QFT is the transform with base $N=2^n$,
corresponding to the n qubit defined by
\begin{equation}
 \ket{x} \to \frac{1}{\sqrt{q}}\sum_{c=0}^{N-1}e^{\frac{2\pi icx}
{N}}\ket{c}.
\end{equation}
Shor proposed the algorithm factoring a composite integer
by the QFT\cite{Shor94}.
The QFT can be constructed by the sequence\cite{Shor94}
in the order (from right to left)
\begin{eqnarray}
&& H_0 C_{0,1}(\theta_1) C_{0,2}(\theta_2) \cdots C_{0,n-1}(\theta_{n-1})
H_1 \cdots H_{n-3}  \nonumber \\
&&  C_{n-3,n-2}(\theta_1)C_{n-3,n-1}(\theta_2) H_{n-2}
C_{n-2, n-1}(\theta_1) H_{n-1}
\label{qft}
\end{eqnarray}
followed by a bit reversal transformation,
where $\theta_j \equiv \pi / 2^j $.
The $n$ qubit QFT
requires $n(n-1)/2$ controlled phase-shifter $ ( C_{j,k}(\theta_{k-j} )$s
and $n$ Hadamard transformation $H_j$s,
and then $n(n+1)/ 2 \simeq O(n^2)$\cite{Shor94}.
This estimation  is based on the complexity analysis method.
It coincides with an actual time-cost 
in the case where all gates required in the sequence(\ref{qft})
are practical elementary ones. 

The controlled phase rotation $C_{j,k}(\theta_{k-j})$ can be
achieved by the sequence
\begin{eqnarray}
C_{j,k}(\theta_{k-j}) &  = & R_{zk}(-\theta_{k-j+1})\Phi_k(\theta_{k-j+2})
R_{zj}(-\theta_{k-j+1}) \nonumber \\
& & XOR(j,k) R_{zj}(\theta_{k-j+1})XOR(j,k). \label{controlled_phase_rotation}
\end{eqnarray}

The intensity control mode  can make
all operations $( C_{j,k}(\theta_{k-j}), H_j$) take almost the same time.
Consequently, the actual time-cost coincides with
the ideal cost in this mode.

The duration control mode, however, makes
each operation $C_{j,k}(\theta_{k-j})$ require  the time  $\tau_{k-j}$ 
in proportion to the phase rotation angle $\theta_{k-j}$.
The operation $C_{0,n-1}(\theta_{n-1})$
rotates the minimum phase angle $\theta_{n-1}$ 
and takes the minimum time  $\tau_{n-1}$
of all rotations in the QFT(\ref{qft}).
The range of required phase rotations in the QFT(\ref{qft})
increases with $2^n$.
For example, 
the time ratio $\tau_{0}/\tau_{n-1}$ approximates to $2^{100} \simeq 10^{30}$
in the 100 qubit QFT.

In general, the time resolution $t_R$ 
controlling the external field is determined by
the response time of the system, the delay of the electronic signal and so on.
We can only execute in the physical implementations 
that satisfy the relationship:
\begin{equation}
\label{resolution}
\tau_{0} >  \tau_{1} > \cdots > \tau_{n-1} \geq t_R.
\end{equation}

It is important for the actual time-cost estimation
to determine how we set up the unit time $t_{unit}$.
The unit time $t_{unit}$ should be also greater than 
the time resolution $t_R$.
The QFT's in these implementations have various actual time-costs
from $t_{unit} = \tau_{0} $ to $ t_{unit} = \tau_{n-1} $.

If we adopt the maximum rotation time $\tau_0$  as the unit time $t_{unit}$,
the actual time-cost is $O(n)$, since
\begin{eqnarray}
\sum_{j=0}^{n-2}\sum_{k=j+1}^{n-1} \frac{\tau_{k-j}}{\tau_{0}}
& = & \sum_{j=0}^{n-2}\sum_{k=j+1}^{n-1} \frac{\theta_{k-j}}{\theta_{0}}
= \sum_{j=0}^{n-2}\sum_{k=j+1}^{n-1} 2^{j-k} \nonumber \\
& = & n + 2^{1-n} -2
\simeq O(n).\label{cost1} 
\end{eqnarray}

On the other hand,
the condition $t_{unit} = \tau_{n-1}$ makes 
the actual time-cost $O(n2^n)$, since
\begin{eqnarray}
\sum_{j=0}^{n-2}\sum_{k=j+1}^{n-1} \frac{\tau_{k-j}}{\tau_{n-1}} 
& = & \sum_{j=0}^{n-2}\sum_{k=j+1}^{n-1} \frac{\theta_{k-j}}{\theta_{n-1}}
\nonumber \\
 & = & \sum_{j=0}^{n-2}\sum_{k=j+1}^{n-1} 2^{n-1+j-k} \nonumber \\
& = & (n-2)2^{n-1} + 1 \simeq O(n2^n).\label{cost2} 
\end{eqnarray}

In this way, the actual time-cost varies from $O(n)$ to $O(n2^n)$, 
and it depends on which of these is adopted as the unit time $t_{unit}$.
The former time-cost however, is  not valid for any $n$
in the following way.

The former condition,  $t_{unit} = \tau_0$, means that
all  phases are always rotated by the external field with constant intensity B
for various data of magnitude $n$.
In this case,
the minimum time $\tau_{n-1}$ decreases exponentially with increasing $n$.
We cannot rotate the phase $\theta_j$ 
to satisfy the condition $\tau_{j} < t_{R}$.
There exists the upper bound $n_{b}$,
satisfying the relationship(\ref{resolution})
for the intensity B under consideration.
The estimated time-cost(\ref{cost1}) is  valid 
for any $n$ satisfying $n \leq n_b$.
It is, however,  not valid for any $n$ which is greater than $n_b$.

The latter condition $t_{unit} = \tau_{n-1}$ means that 
the intensity B decreases exponentially with  $n$
satisfying the relation(\ref{resolution}).
In this case, 
we can rotate all phase angles $\theta_{j} ( 0 \leq j \leq n-1 ) $
in the QFT(\ref{qft}) accurately.
A particular condition $t_{unit} = \tau_{n-1} = t_R$ always  achieves
all phase rotations in the minimum total time,
and then yields the best computing performance for each value of  $n$. 
The actual time-cost always obeys eq.(\ref{cost2}) for any $n$.

In this way, 
the actual time-cost varies from $O(n)$ to $O(n2^n)$ for any $n(\leq n_b)$,
and follows only the latter for all other values of $n$.
These costs are estimated making  the assumption that
the QFT is always executed accurately.

An approximate QFT, (AQFT) can reduce the arbitrary numbers of
the controlled phase shift gates by sacrificing the accuracy\cite{Barenco96}.
We can select the AQFT with an actual time-cost 
between $O(n)$ and $O(n2^n)$, 
by considering the required accuracy for any $n$.

We have discussed the actual time-cost 
from  the viewpoint of the phase rotations in the QFT(\ref{qft}).
Almost $C_{j,k}(\theta_{k-j})$ operations 
occur for  non-adjacent registers in the QFT.
We need to construct such non-adjacent gates using adjacent ones 
so here we estimate the actual time-cost required 
for constructing such non-adjacent gates from adjacent ones. 
We must construct any non-adjacent gates 
by use of the adjacent swap technique.
The swap $S_{j,k}$ is an operation for
exchanging data between two quantum registers simply.
It is achieved via the sequence
$S_{j,k} =XOR(k,j) XOR(j,k) XOR(k,j)$.
The non-adjacent two qubit gate $U_{j,k}$
is achieved by adjacent swaps and adjacent $U_{l,l+1}$
through the following $k-j-1$ sequences.
\begin{eqnarray}
U_{j,k}
&= &S_{k,k-1}S_{k-1,k-2}\cdots S_{j+3,j+2}S_{j+2,j+1} U_{j,j+1} \nonumber \\
& & S_{j+1,j+2} S_{j+2,j+3} \cdots S_{k-2,k-1} S_{k-1,k} \label{non_adj_1} \\
&=& S_{k,k-1}\cdots S_{l+2,l+1}  S_{j,j+1} \cdots S_{l-1,l}U_{l,l+1}\nonumber \\
& & S_{l,l-1} \cdots S_{j+1,j} S_{l+1,l+2}\cdots S_{k-1,k} 
( j < l < k ) \nonumber  \\
&=& S_{j,j+1} S_{j+1,j+2} S_{j+2,j+3} \cdots S_{k-2,k-1} U_{k-1,k}\nonumber \\
& & S_{k-1,k-2} \cdots S_{j+3,j+2}S_{j+2,j+1}S_{j+1,j}.\label{non_adj_2}
\end{eqnarray}
Figure 1 represents the sequence(\ref{non_adj_1}).
These sequences show that
the non-adjacent operation $U_{j,k}$ requires
both the time for the adjacent operation $U_{j,j+1}$
and another time in proportion to $|k-j|$ for swaps transferring data.

Considering the construction (\ref{non_adj_1}$\sim$\ref{non_adj_2}),
the QFT(\ref{qft}) requires 
more  $(n-1)n(2n-1)/6 \sim O(n^3)$ adjacent swaps for data transfer
besides the adjacent controlled phase rotations.

There is, however, a way to reduce some swaps.
From sequences(\ref{qft}) and (\ref{non_adj_2}),
we can obtain the actual quantum circuit shown by Fig.2.
In the figure, swaps $S_{j,j+1}\cdots S_{k-1,k}S_{k,k-1} \cdots S_{j+1,j}$
inside each box can be reduced to the identity operation$I$.
The $n$ qubit QFT requires $(n-1)(n-2) \sim O(n^2)$ swaps as a result.
It has the same polynomial order as  the ideal cost.
In this way,
there are some cases where the swaps can be reduced 
due to the symmetry of the algorithm under consideration.

We can conclude that 
the actual time-cost of the QFT is dominated by the phase rotations
and then are $O(n2^n)$  in the range $n > n_{b}$.

The range of required phase rotation angles
increases exponentially with $n$ in the QFT if the accuracy is preserved.
In the implementations that are considering,
we need to obtain it by controlling the duration or the intensity
of the external field.
The duration control mode
increases the actual time-cost drastically with $2^n$.
The intensity control mode in BSR
takes least time and seems to be the most efficient case for the QFT.
This mode, however, leads to  another burden on the equipment.
The required range of intensity for the applied field
increases with $2^n$.
If the minimum phase rotation is implemented by the field $B=10^{-3}$ T,
the maximum phase rotation for the 100 qubit QFT
requires a field intensity $B \sim 10^{27}$ T,
which is far beyond the current feasible intensity 
for the magnetic field.

Our results lead to concerns about 
the feasibility of factoring huge numbers 
with polynomial order time-costs in the implementations we are considering.

We have shown that almost all universal gates do not expend the same time
in BSR and the SRT.
This causes the actual time-cost of the QFT to be  drastically
different from the ideal one.
The ideal cost of the QFT is not effective in the range  $n > n_{b}$
 for practical cases of these implementations 
 if the accuracy is preserved.

We believe that both ideal and actual discussions are
important for the development of quantum computer science.
The former discussion stimulates  study of 
the characteristic of the algorithm itself,
and the latter is also important for the current situation
in  quantum computation as we move 
from the conceptual system to the physical one.
Our discussion shows the necessity
of discussing the practical elementary gate
in other proposed quantum computers,
and of estimating the actual time-cost for other quantum algorithms.

\setlength{\unitlength}{0.015in}
\begin{figure}
\begin{picture}(30,90)(-10,0)
\bla{5}{0}{15}
\ml{0}{30}{20}{2}
\ml{0}{75}{20}{2}
\bla{5}{15}{15}
\put(5,58){\vdots}
\put(15,58){\vdots}
\put(10,90){\circle*{2}}
\put(10,90){\line(0,-1){70}}
\put(5,10){\framebox(10,10){U}}
\end{picture}
\begin{picture}(20,100)(0,0)
\put(10,40){\makebox(10,10){$=$}}
\end{picture}
\begin{picture}(130,90)(0,0)
\ml{0}{30}{120}{2}
\ml{0}{75}{120}{2}
\bla{55}{0}{15}
\bla{55}{65}{15}
\put(5,70){\framebox(10,25)}
\put(10,75){\swap}
\put(20,60){\swap}
\put(30,58){$\vdots$}
\put(40,45){\swap}
\put(50,30){\swap}
\put(60,30){\circle*{2}}
\put(60,30){\line(0,-1){10}}
\put(55,10){\framebox(10,10){U}}
\put(70,30){\swap}
\put(80,45){\swap}
\put(90,58){$\vdots$}
\put(100,60){\swap}
\put(110,75){\swap}
\end{picture}
\caption{Non-adjacent operation  achieved by
utilising  adjacent operations.
The symbol inside the box represents the adjacent swap $S_{j,k}$.}
\end{figure}
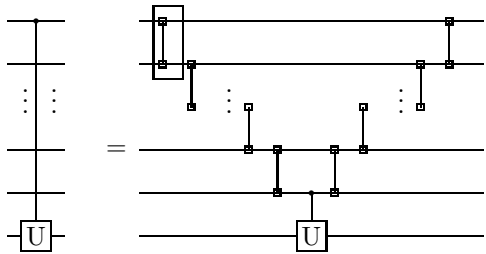

\begin{figure}
\begin{picture}(200,85)(0,0)
\numa
\ml{0}{15}{10}{5}
\ml{10}{15}{50}{4}
\bla{10}{20}{75}
\bla{160}{40}{75}
\bla{130}{70}{60}
\ml{60}{15}{50}{3}
\ml{110}{15}{10}{2}
\ml{120}{15}{60}{3}
\bla{10}{180}{15}
\bla{10}{180}{45}
\ml{190}{15}{10}{3}
\h{10}{70}

\c{30}{70}
\put(35,60){\line(0,1){10}}
\put(35,60){\circle*{2}}
\put(50,60){\swap}
\c{60}{55}
\put(65,45){\line(0,1){10}}
\put(65,45){\circle*{2}}
\put(80,60){\swap}

\put(75,55){\framebox(20,25)}

\put(90,60){\swap}
\put(100,45){\swap}
\c{110}{40}
\put(115,30){\line(0,1){10}}
\put(115,30){\circle*{2}}
\put(130,45){\swap}
\put(140,60){\swap}

\put(125,40){\framebox(40,40)}

\put(150,60){\swap}
\put(160,45){\swap}
\put(170,30){\swap}
\c{180}{25}
\put(185,15){\line(0,1){10}}
\put(185,15){\circle*{2}}
\put(207,0){\numb}
\put(100,5){\makebox{(a)}}
\end{picture}

\begin{picture}(200,85)(0,0)
\numb
\bla{200}{0}{75}
\ml{0}{15}{90}{3}
\bla{40}{0}{60}
\bla{10}{50}{60}
\bla{130}{70}{60}
\ml{90}{15}{10}{2}
\ml{100}{15}{40}{3}
\bla{10}{140}{15}
\bla{10}{140}{45}
\ml{150}{15}{50}{3}

\put(10,30){\swap}
\put(20,45){\swap}
\put(30,60){\swap}

\h{40}{55}

\c{60}{55}
\put(65,45){\line(0,1){10}}
\put(65,45){\circle*{2}}

\put(80,45){\swap}
\c{90}{40}
\put(95,30){\line(0,1){10}}
\put(95,30){\circle*{2}}
\put(110,45){\swap}

\put(105,40){\framebox(20,25)}

\put(120,45){\swap}
\put(130,30){\swap}
\c{140}{25}
\put(145,15){\line(0,1){10}}
\put(145,15){\circle*{2}}
\put(160,30){\swap}
\put(170,45){\swap}
\put(207,0){\numc}

\put(100,5){\makebox{(b)}}
\end{picture}

\begin{picture}(200,85)(0,0)
\numc
\ml{0}{60}{200}{2}
\bla{10}{0}{45}
\bla{10}{20}{45}
\bla{160}{40}{45}
\bla{60}{0}{30}
\bla{20}{70}{30}
\bla{10}{100}{30}
\bla{80}{120}{30}
\bla{130}{0}{15}
\bla{60}{140}{15}

\h{10}{40}
\c{30}{40}
\put(35,30){\line(0,1){10}}
\put(35,30){\circle*{2}}
\put(50,30){\swap}
\c{60}{25}
\put(65,15){\line(0,1){10}}
\put(65,15){\circle*{2}}
\put(80,30){\swap}

\h{90}{25}

\c{110}{25}
\put(115,15){\line(0,1){10}}
\put(115,15){\circle*{2}}

\h{130}{10}

\put(100,5){\makebox{(c)}}
\end{picture}

\caption{Figures (a),(b) and (c) in order are
the actual quantum circuits of the 5 qubit QFT.
This quantum circuit is derived from sequences 
(4) and (10).
The swaps inside the boxes are reducible. }
\end{figure}
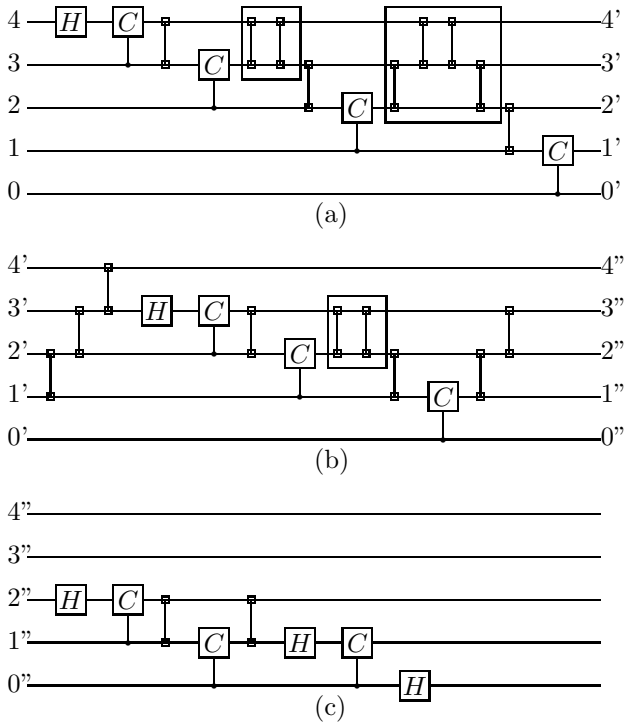

\bibliographystyle{unsrt}

\end{document}